# Broadband super-resolution Terahertz Time domain spectroscopy applied to Gas analysis

S. Eliet,[1] A. Cuisset,[2,3] F. Hindle,[2,3] J.F. Lampin,[1] and R. Peretti[1*]

[1]Institut d'Électronique de Microélectronique et Nanotechnologies, UMR CNRS 8520 Villeneuve d'Ascq, France

[2]Univ. Littoral Côte d'Opale, UR 4493, LPCA, Laboratoire de Physico-Chimie de l'Atmosphère,

[3] F-59140 Dunkerque, France

* Corresponding author: *Romain.peretti@univ-lille.fr*

**Abstract:** Terahertz (THz) Time domain spectroscopy (THz-TDS) is a broadband spectroscopic technique spreading its uses in multiple fields: in science from material science to biology, in industry where it measures the thickness of a paint layer during the painting operation. Using such practical commercial apparatus with broad spectrum for gas spectroscopy could be a major asset for air quality monitoring and tracking of atmospheric composition. However, gas spectroscopy needs high resolution and the usual approach in THz-TDS, where the recorded time trace is Fourier transform, suffers from resolution limitation due to the size of the delay line in the system. In this letter, we introduce the concept of constraint reconstruction for super-resolution spectroscopy based on the modeling of the spectroscopic lines in a sparse spectrum. Light molecule gas typically shows sparse and narrow lines on a broad spectrum and we propose an algorithm reconstructing these lines with a resolution improvement of 10 the ultimate resolution reachable by the apparatus. We envision the proposed technique to lead to broadband, selective, rapid and cheap gas monitoring applications.

Terahertz (THz) Time domain spectroscopy (THz-TDS) began its development with the birth of the femtosecond laser as sub-picosecond sampling [1]. The idea was quickly used for spectroscopy [2] and few years later, the integration of photoconductive antenna [3, 4] bring it to a practical broadband THz spectroscopic system. These systems are now commercial and offer more than one decade of bandwidth. They enabled myriads of applications from semiconductor spectroscopy [5, 6], 2D materials [7] or biological samples [8]. Results about THz-TDS gas spectroscopy are rarer. Still, such broad bandwidth up to 6 THz and more would enable fast multi-species gas analysis. Due to its ability to non-invasively probe innumerable chemical species in a single spectrum, it would be used for instance for air quality monitoring and tracking of atmospheric composition. In fact, this has been tried as early as in 1989 [9] but the community quickly realized that the length of the delay line, leading the time sampling was a limitation to its use. In THz-TDS, the signal is recorded only on a limited time window limiting the resolution when performing the Discreet Fourier Transform (DFT) to a limitation that we will call the Fourier criteria here (*Fc*). DFT projects the signal on a basis of functions that is a frequency comb. The spacing between the teeth of the comb is equal to the inverse of the time windows length is the resolution limitation *Fc*. For gas spectroscopy and sensing applications, the needs for specificity and selectivity require to reduce the collisional broadening up to the Doppler linewidth limit which is reached at low pressure (<10 mbar). In this case, the apparatus face the challenge of narrow and sparse, fully or partially, resolved rotational lines requiring high-resolution techniques [10, 11]. In THz-TDS, one idea could be to use very long delay lines. However, this would terribly degrade the signal to noise ratio, the stability and the compactness of the systems. Other experimental approaches based on the use of optical frequency combs enabled new broadband spectroscopic setups of particular relevance for THz molecular spectroscopy [12]. However, these systems are still, extremely complex to operate, costly and the experiments last hours if not days. Signal processing attempts were made [13] to unravel THz-TDS potential for high-resolution measurements. In fact, in other fields, as quantum chemistry [14], nuclear magnetic resonance [15], or photonic simulations [16] methods, such as wavelet transform [17], filter diagonalization [18] or harmonic inversion [14] allowed resolution performance improvements far beyond $Fc$ because they do not use basis functions without pre-fixed central frequency. We propose here to implement such methods to enable THz-TDS super resolution spectroscopy.

On Fourier transform interferometers, super-resolution spectroscopy was already tried from using maximum entropy methods [19] and compact representation [20]. Recently, a new set up based on a random laser and a two-path interferometers opened new approach for practical super resolution spectroscopy [21] and another examples raised for Raman spectroscopy [22]. In this work, while the methods is valid for any spectroscopy technique involving a frequency comb, we implemented a Constrained Reconstruction Super Resolution (CRSR) algorithm leading to super-resolution spectroscopy on a commercial broadband (~0.2 – 6 THz) THz Time Domain Spectrometer with a Fourier transform limited spectral resolution of $Fc$ = 1.2 GHz. We validated the methods on rotational spectra of gas phase ammonia ($NH_3$) with the measurements of Doppler limited and self-broadened rotational lines in the 1 – 700 mbar pressure range.

A major difference between spectral resolution in spectroscopy and spatial resolution in microscopy is the shape of the detected object. A microscope target is of an arbitrary shape but the profile of a spectroscopic line follow quantum mechanics and thus is much more constrained. This "*a priori*" knowledge is very important since it allows the following of a constrained reconstruction approach [23] without any information loss. In other words, in the frequency domain, rovibrational lines of molecules diluted in a gas sample follow either a Gaussian or a Lorentzian profiles, corresponding respectively to the Doppler and collisional broadenings. At the pressures considered in this study, the collisional effects dominate leading to a simple Lorentzian profile. This corresponds, in the time domain, to an exponentially damped sinus function. The super-resolution relies on the fact that each damped sinus are described by only three parameters: the amplitude, the central frequency and the damping rate. As seen on figure 1, one needs only few time domain experimental points to retrieve the parameters. Still, raises the limitation of having fewer oscillators than a fraction of the total number of points in the full time trace (this limitation is called generalized Fourier-Heisenberg uncertainty [14]). With this constraint, the resolution will only be limited by the experimental signal-to-noise ratio and the computational methods to optimize the fit.

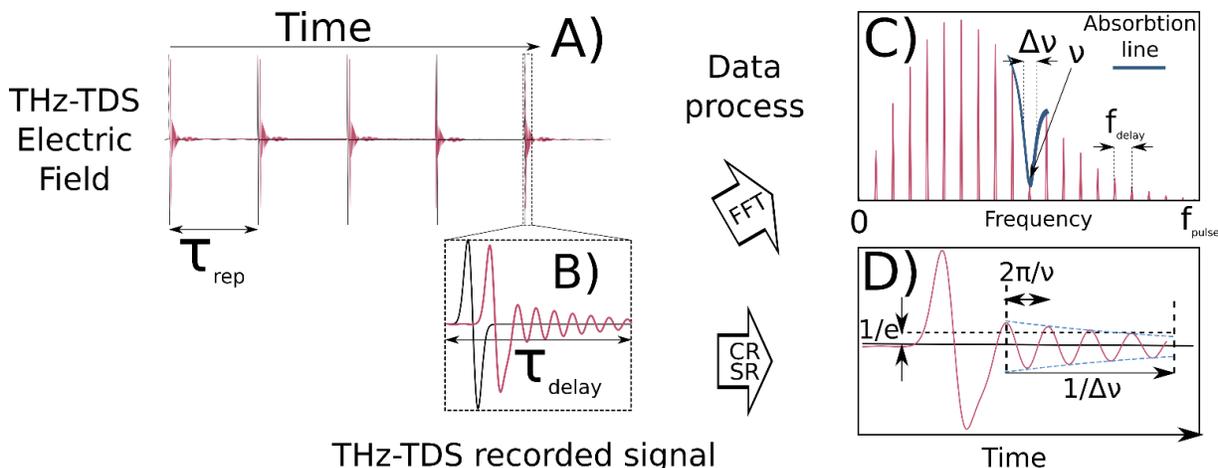

FIG. 1. Principle of the methods. In THz-TDS the signal recorded in the time domain (A) is limited by the time windowing coming from the travel range of the delay line (see zoomed part in B). Usually, a discreet Fourier transform is applied to this signal (C). The CRSR in (D) uses the same data and perform a full analysis in the time domain to retrieve the parameters of each spectral lines, which correspond in the time domain to the damped sinus (bottom right) and thus achieve super resolution.

To retrieve the three Lorentz parameters, namely the central frequency $\nu$, the linewidth $\Delta\nu$ and the intensity $\Delta\varepsilon$[1], we implemented the super-resolution in an open source software fit@TDS previously developed by our group [24, 25]. We coded the software based on a method where we used a multi-parameters nonlinear optimization algorithm to retrieve parameters from THz-TDS. The super –resolution approach surpass the previous version because it uses a different time-frame. Previously, we used the time-frame of the recorded signal: $\tau_{delay}$ in figure 1 making the assumption that the signal was lower than the noise at the end of the time trace [25]. Here, we are using the real period of the experiment i.e. the repetition time interval of the excitation laser: $\tau_{rep}$ in figure 1. As such, DFT in the model and experiments follow the same periodicity preventing any artificial DFT-folding artifact or aliasing and reproducing the ones from the experiments. Precisely, a line narrower than the Fc limitation will give rise to a temporal signal going over the right edge of the time window and coming back on the beginning of it on the left edge as explained in [12]. Since the length of the delay line does not allow the recording of the full time-frame, we compared only the recorded fraction to the model. We performed an inverse Fast Fourier Transform (*FFT*) to get modeled data in the time domain at each step and calculated the residual error by taking the root mean square difference to the experimental data. We used this residual error in the optimization problem set in either an augmented *Lagrangian* particle swarm [26] or a sequential least squares [27] optimizer.

We assessed the CRSR by performing absorption measurements of $NH_3$ using a commercial transportable all-fiber high-end THz-TDS system Terasmart from Menlo Systems GmbH. To suppress the Fabry-Perot echoes we set a Brewster-angle Silicon-windows gas-cell, with an optical path length of 7 cm, in the collimated THz beam. The cell was evacuated with a turbomolecular pump and pure $NH_3$ (99.98%) was injected in the cell to obtain the targeted pressure. From 1 to 200 mbar a thermostated Baratron® capacitive gauge was used with a precision of 0.12%. From 200 mbar to atmospheric pressure, an Edwards CG 16K capsule dial gauges barometrically compensated was used with a precision of 3 %. We repeated the thirteen minutes long experiments (1000 acquisitions of 780 ms) for pressures from 1 mbar to 690 mbar. The time window was 848 ps wide (time sampling of 33 fs) leading to a $Fc$ of 1.2 GHz. Thanks to this set up we recorded the time traces given as supportive information. We applied the CRSR on the recorded time traces and assed it regarding three key performances. First its ability to point the center of a line. Second its ability to determine its linewidth. Finally, its ability to distinguish two close spectral lines. The results are depicted in figure 2.

---

[1] $\Delta\varepsilon$ is in dielectric constant unit and is related to the intensity line parameter $S$ by $\Delta\varepsilon = \frac{2cSP}{k_b T(2\pi\nu)^2}$, with $c$ the celerity, $P$ the pressure, $k_b$ the *Boltzmann* constant and $T$ the temperature.

Panel A shows the reconstructed spectrum of the isolated absorption rotational line around 572.5 GHz for three pressures (6, 55 & 580 mbar) compared to the FFT of the time trace. For the widest line measured at P=580 mbar, the CRSR gives similar results to the FFT. For the line at 55 mbar and the one at 6 mbar, the FFT principally exhibits a modification of a single frequency point. This means that for the line center we can only be state that it is 572.5 ± 1.2 GHz and for the linewidth, that it is below 1.2 GHz, Instead, CRSR gives a center of 572.52 and 572.72 GHz and a linewidth of 1 118 MHz and 114 MHz respectively, for these two lines. To evaluate the accuracy of this reconstruction, we plotted on panel B and C the variation of the extracted line-frequency and linewidth with the gas pressure and we compared these two line parameters with the values taken from the self-shift and the self-broadening coefficients tabulated in the Hitran database [28, 29].

The first super-resolution criteria: the central frequency corresponds to the database value at a precision of 20 MHz down to a pressure of 3 mbar (relative precision: 3.5 $10^{-5}$ at 573 GHz). This precision exceeds the one of the commercial continuous wave systems [30] and is even a measurement of the absolute central frequency thanks to the calibration done on water vapor lines (see supportive information -5). We attribute the major contribution to the line frequency error to the pressure measurement uncertainties for the highest pressures and to the decreasing of the SNR at lowest pressures.

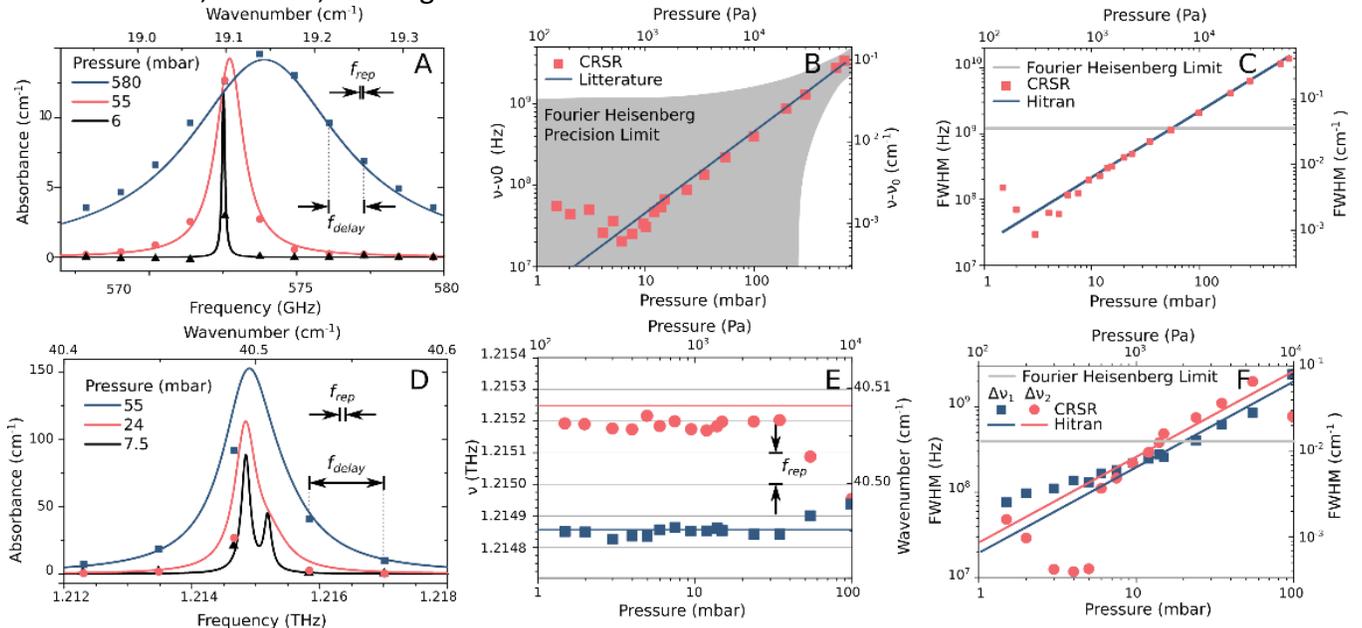

Fig. 2. Super resolution in the $NH_3$. Panel A, shows the reconstructed 573 GHz lines in solid line superimposed to the usual FFT spectrum (symbols). On panel B, the frequency shift regarding the tabulated HITRAN value [28] of the line frequency at null pressure is compared to the one retrieved by CRSR. The gray area depicts precision limitation of the Fourier limitation (it is calculated by adding, for the upper bound, and subtracting, for the lower bound, the $Fc$=1.2 GHz to the theoretical value of the literature). The panel C shows the linewidth versus the pressure. Panel D and F are similar to panel A and C

respectively for the doublet of $NH_3$ rotational lines around 1.21 THz. Panel E shows the central frequency of the two lines of the doublet versus pressure compare to the literature value.

The second criteria: the linewidth corresponds to the one of the database better than 5 % from atmospheric pressure down to 10 mbar before a progressive increasing of the error at lower pressures. We retrieved an accurate linewidth down to a pressure of 6 mbar corresponding to a linewidth of 114 MHz with a precision of ~15 MHz. This corresponds to a 10-fold improvement super resolution compared to the *Fc*.

To reach the third criteria and achieve the doublet separation, we focused on how to discriminate two absorption lines close together, namely the 1.21 THz $NH_3$ rotational line doublet separated by ~393 MHz. Here, we aim at retrieving twice the number of parameters. For this reason, we added an additional constraint to the CRSR. For the first and only time of the study we used HITRAN database as a partial input and we fixed the ratio of oscillator strength between the two lines to the literature value of 0.38 [28]. On the Panel D of figure 2, one sees the doublet at 55 mbar is very similar to a single line. Still, we depicted it as the sum of two Lorentz oscillator from which parameters are plotted on panel E and F, where one can see the very good agreement with the theoretical values. At 24 mbar when the shape of the line begins to show a shoulder, and at 7.5 mbar when the two lines are separated and mainly modify the absorption on a single spectral point, the retrieval is very good. Once again, to check the reliability of the CRSR, we compare the retrieved parameter to the HITRAN tabulated ones used as reference. On panel E, from the comparison of the central frequencies of both lines of the doublet at very low pressure, we estimate the absolute pointing precision to ~50 MHz, namely a relative precision of $4\ 10^{-5}$ at 1.215 THz. Moreover on panel E, the comparison shows two limits on the line frequencies determination: at very low pressure (< 7 mbar) the SNR and at high pressure (> 80 mbar) the overlapping of the line as demonstrated by the numerical example in SI (fig S1 and 1§). Still, the values are very accurate from a pressure of 7.5 mbar up to 55 mbar. In this range, the relative accuracy is better than 30 % for a line as narrow as 150 MHz. This demonstrates that CRSR reach the third criteria for super resolution: we succeeded in reconstructing the two lines of a doublet separated by less than the usual resolution of the apparatus.

In conclusion, we implemented a super-resolution algorithm for THz-TDS and tested it on a broadband sparse gas phase spectrum exploiting the wealth of information lying in the THz-TDS time traces. This was achieved on rotational transitions of $NH_3$ on a spectrum spanning more than a decade (36 lines from 200 GHz to 4 THz with a strength line between $1.10^{-20}$ and $4.610^{-19}$ $cm^{-1}/molecule.cm^{-2}$) with more than 10 super-resolution retrieved lines from 570 GHz up to 3.5 THz (Fig S3 & table S1). Our method delivers accurate reconstruction despite minimal algorithm optimization and the moderate experimental integration time leading to 13 minutes experiments compared to days one when using FTIR with the same resolution and number of scan [31]. We have shown a spectral enhancement factor of above three for doublet separation, above 10 for the linewidth measurements and the absolute line center pointing. These limitations come from the presence of noise in the experiments and are thus limited by the system used. In absolute terms, an ultimate resolution limit is represented by the linewidth of the repetition rate of the laser frequency comb (which, in our case, was around 30 Hz at 100 MHz).

In principle, the methods can be adapted to any spectroscopy technics in the reciprocal domain such as THz-TDS or Michelson based Fourier transform spectroscopy, but would be even more efficient for spectroscopic methods based on frequency combs as dual comb spectroscopy.

This is specifically salient for gas from the THz to the IR range where a high resolution is needed.

Most importantly, our approach relies entirely on the sparse nature of the acquired spectrum, and was achieve on a commercial spectrometer without active stabilization of the laser. Moreover, THz-TDS technics are making progress every year because of improvements of the THz photoconductive antennas [32], the laser source used, the delay lines and the system itself. Nowadays several systems allow time sampling without mechanical delay lines using dual comb spectroscopy in asynchronous optical sampling [33]. All these improvements will increase both the bandwidth of the spectrum and the signal to noise ratio. The direct consequence will be a better super resolution enhancement factor on a broader spectrum and will soon enable to reach a resolution closer to the Doppler limit.

Finally, the strong rotational lines of a large variety of polar molecules lie in the THz. Therefore, THz spectroscopy find applications from fundamental sciences specifically in astronomy (astrophysics, planetology), as well as in daily subjects in atmosphere monitoring regarding pollution and global warming or gas analysis for health purposes. The rapid analysis broadband of numerous gases requires over several THz and with a subMHz resolution for reliable species discrimination and quantification. The CRSR algorithm will provide a resolution better than $f_{rep}$.with a simple tabletop, convenient, commercial THz apparatus and we anticipate that it will be the key to the expansion of THz-TDS of gas.

**Data Availability Statement**

Most of data have been added as supportive information and the authors will make further data available under reasonable request.

# Supplementary document

**1. Benchmark : Numerical validation**

For benchmarking purposes, we modeled the absorption of several pressures of $^{14}NH_3$ in the THz range with an optical path length of 7 cm as in the experiments. Here, $f_{rep}=1/\tau_{rep}=100$ MHz and $f_{delay}=1/\tau_{delay}=1.178$ GHz. Similarly, to the experiments, we focused on three key performances. First its ability to point the center of a line. Second its ability to determine its linewidth. Finally, its ability to distinguish two close spectral lines. The intense line around 572.5 GHz of the J=0 →1, K=0 rotational transition is used to test our two first criteria and the doublet J=1 →2, K=0 & K=1 around 1.21 THz for the last one. We convoluted an experimental reference with the model of the lines using the parameters found in the HITRAN database [28] and we added a Gaussian white noise on the resulting time traces of same magnitude as in the experiments corresponding to a dynamic range of 70 dB. We then used the CRSR on these data. The results are depicted in fig 2.

There, we compare the targeted ideal line from HITRAN to the retrieved one computed using code file 2 and given in data file 1. To figure out the benefit of CRSR, we added on the same plot the usual spectrum plotted in TDS in large symbols (the sampling corresponds to the $\tau_{delay}$ time window) and the same data if we could record the time trace with a time window $\tau_{rep}$.

For the case of the isolated line at 572.5 GHz, the broad absorption feature is resolved at 60 mbar by only three $f_{delay}$ datapoints and many $f_{rep}$ points. The linewidth here represents the resolution limit for a standard TDS instrument using a $\tau_{delay}$ time window. At 6.0 mbar the line is some ten times narrower than $f_{delay}$ and of similar width compared to $f_{rep}$. With a further reduction in pressure to 0.6 mbar the linewidth is one hundred times narrower than $f_{delay}$ and ten times narrower than $f_{rep}$. At these two latter pressures, the line cannot be correctly resolved using the standard approach taking the Fourier transform the temporal recording.

The CRSR algorithm accurately retrieves the real line parameters in all the cases including at the lowest pressure. Both the central frequency (including the pressure shift) and the linewidth are retrieved very precisely, validating our two first assessment objectives. On panel B, the linewidths fall below the usual resolution limit for pressures below $P_{f_{delay}}$. The reconstruction stays accurate above 0.6 mbar. This corresponds to lines much narrower than the resolution limit and even narrower than $f_{rep}$. At pressures below 0.6 mbar, the error becomes too high and the line is not correctly reconstructed. In fact, one can derive the total absorbed energy by the line versus the pressure $P$:

$$SNR \propto \sigma = P \times \frac{P}{P+P_{f_{delay}}} \quad (1)$$

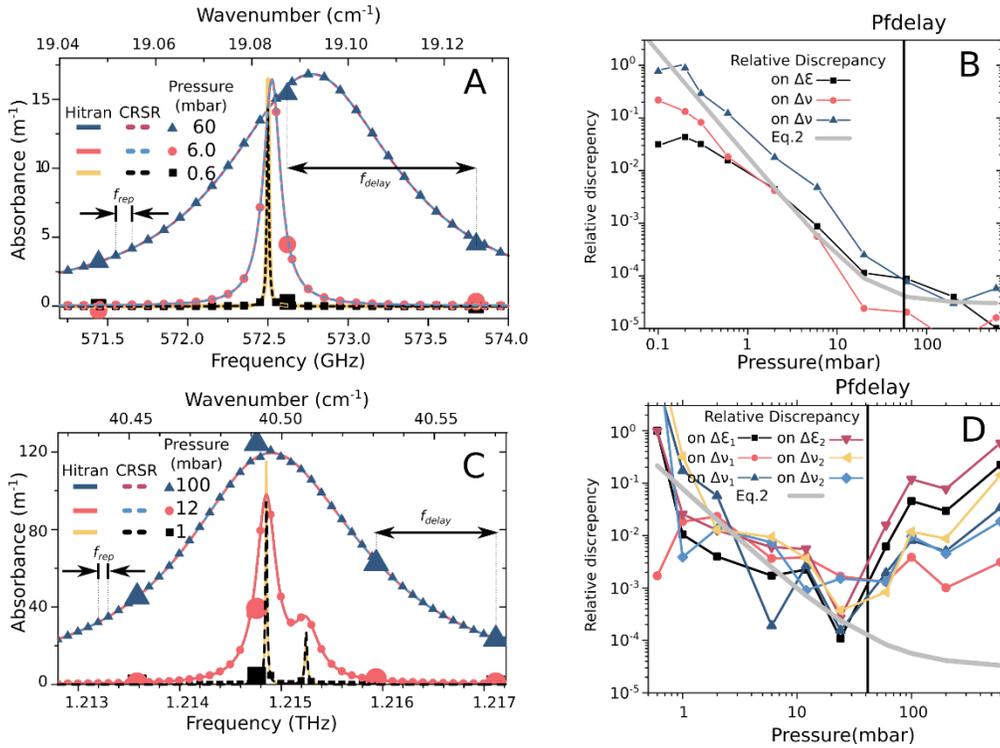

Fig S. 1. Numerical example of Super resolution: results and performances assessment. In panel A the THz-TDS spectrum of the line around 573 GHz, (large symbols) shows how the line spread spectrally using the classical mechanical delay stage. In small symbols, we plotted the FFT of a hypothetical signal where the full time ($\tau_{rep}$ time window) trace would be recorded. The reconstructed super resolution lines (dotted lines) superimposed very well with the targeted line (plain lines). In panel B, the precision of the retrieval of each parameter of the lines is plotted. It demonstrates the very good performance of the CRSR down to very low pressure. Panel C and D are similar, respectively, to panel A and B but for the doublet around 1.21 THz. Panel C demonstrates the ability of CRSR to discriminate a doublet separated by one third of fdelay (393 MHz). On Panel D the comparison with eq. 2 indicates that at low pressure the retrieval is limited by the signal to noise ratio when at higher pressure (when the width of the lines of the doublet are higher than the frequency separation), overlap effects deteriorate the precision of the method.

The first term comes from the number of molecules crossed by the THz beam. The second term comes from the finites size of the delay line and plays a role for lines narrower than $f_{delay}$ (being at $P_{f_{delay}}$). In this case, the time signal extends further than the delay line windows and this fraction of the signal is lost. To analyze this more quantitatively we plotted the floored inverse function of the $SNR$ on figure S1 panel B:

$$E_{rr} = \theta_{noise} + \frac{S_{tot}}{\sigma} \quad (2)$$

Where $S_{tot} = 3\ 10^{-4}$ corresponds to a fitted normalization factor to the signal (set one for all for one transition) and $\theta_{noise}$ (=3 $10^{-5}$, fitted as well) is the best accuracy we can reach considering the noise and the performances of the algorithm. There, the discrepancy on each parameter and the mean square error on the spectrum follows the trend of the equation only coming from the energy absorbed by the line. To conclude, we enable super resolution on lines down to about one tenth of $f_{rep}$ and one hundredth of the usual resolution limitation for this level of noise for the two first criteria. We establish that the resolution performance of the CRSR technique is dependent on the SNR. Meaning the SNR of the molecular signal not of the source. The natural conclusion is therefore for a stronger absorber, or a longer interaction length a resolution improvement may be expected.

The most convincing proof of super resolution is a techniques capacity to resolve doublet. We examined the doublet at 1.21 THz, which exhibits a frequency difference of ~ 393 MHz and stands above $f_{rep}$ and below $f_{delay}$ making it a very good example. It is important to note in performing the usual FFT both lines of the doublet stand in the same frequency point as seen on panel C of fig 2. As in the single line case, we plotted the retrieved spectrum and the error on panel D. Once again, CRSR retrieves very accurately the parameters from 3 mbar to 100 mbar. However, the pressure range of reliability is narrower. At pressures below 1 mbar the doublet is not retrieved properly. More quantitatively, with equation 2 added to panel D with the same $\theta_{noise}$ and a $S_{tot} = 1/629$; to take into account the lower signal at this frequency (~8 dB compared with 572 GHz). As expected, equation 2 explains the low-pressure behavior but does not take into account the effect the overlapping lines occurring above 100 mbar. Here the lifetime of the oscillator (inverse of the linewidth) became shorter than the beating period of the doublet. Thus, the overlap degrades the accuracy of the fit. Still, we enable super resolution of a doublet down to one third of the usual resolution limitation for this level of noise. We want to emphasize here that contrary to what we did for the experimental data, we did not need here the additional constraint of fixing the ratio between the two lines intensity. This shows that the doublet separation is fully achievable with CRSR even without the need of additional constraint as soon as the signal to noise ratio is good enough.

Using fictitious samples of a single line and a doublet, at pressures from 0.1 mbar to atmospheric pressure, including an experimental level of white Gaussian noise, we showed the capacity of CRSR to retrieve the spectroscopic parameters of rotational lines from TDS time traces. With decreasing pressure the signal is reduced, nevertheless CRSR provides accurate parameters down to 0.6 mbar for a single line corresponding to a line narrower than 30 MHz. the resolution demonstrated by this approach is easily sufficient to fully resolve the lines separated by 400 MHz at 1 mbar. This is coherent with the resolution limit observed for a single line of approximately 30 MHz. This means a super-resolution of a 30-fold factor improvement.

2. **FULL spectrum NH3**

To give a better insight on the studied spectrum, Figure S shows the recorded experimental spectrum of 10 mbar NH3 gas overall the TDS bandwidth using FFT on the time data.

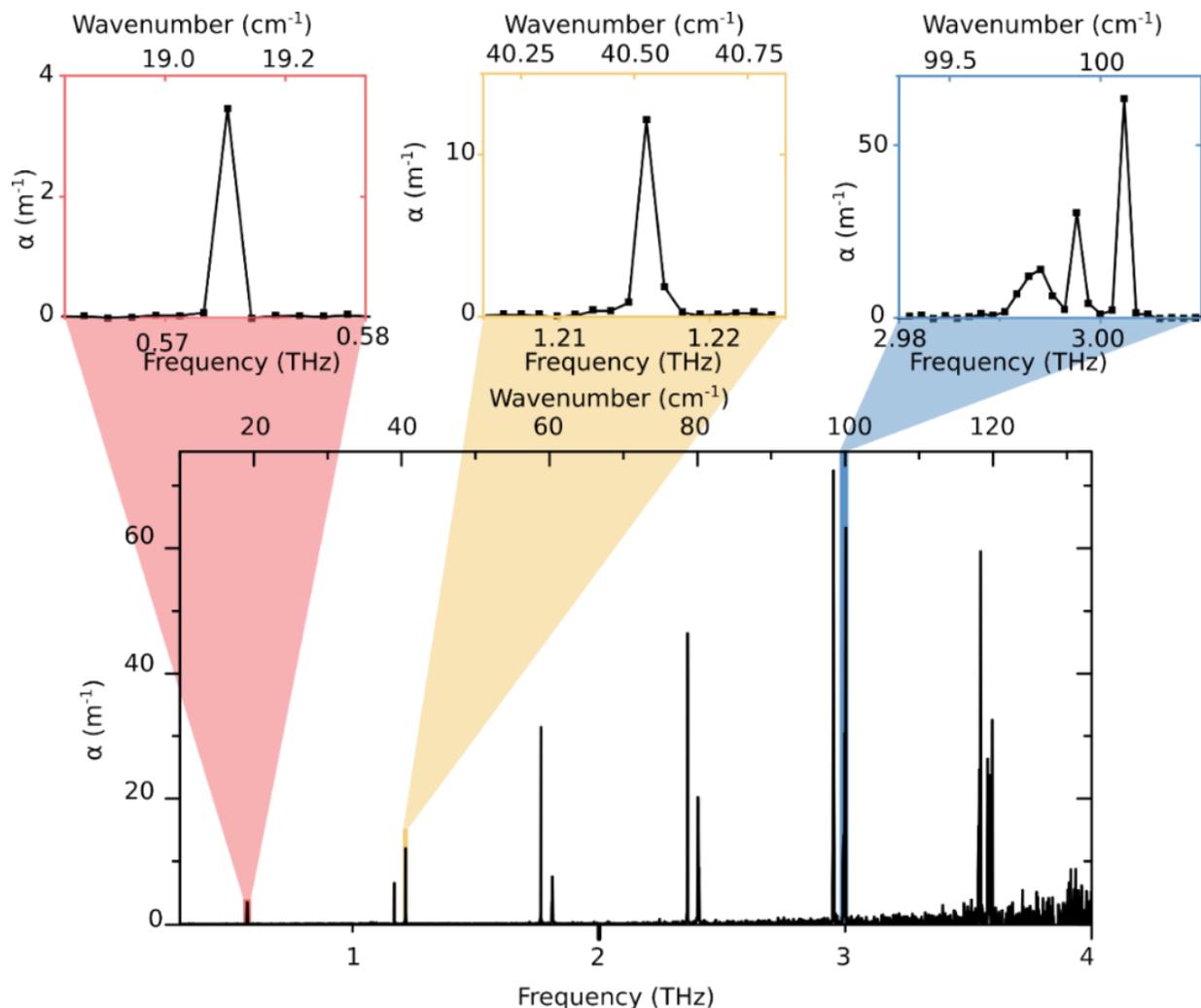

Figure S 2: Experimental spectrum of the absorbance of 10 mbar gaseous NH3 from our experiments (simple FFT) showing that the spectrum extends from 200 GHz to 4 THz with narrow lines distributed from 500 z to 3.6 THz. It was recorded using the 7 cm long Brewster angle gas cell in the TDS setup. The three top insets shows a zoom on the lines studied in the paper.

Here one sees the 36 lines spreading over the broad THz-TDS spectrum (~0.5 to > 3.5 THz). Each of these lines mostly modify one spectral point, but still shoulder effect is appreciable on the surrounding points. Additionally, we want to insist that this assertion remains valid for the doublet at 1.21 THz, meaning that the doublet mostly modify one spectral point.

### 3. Full spectrum reconstruction at 9.5 mbar

In the main text, we focused on individual peaks/doublet to analyze the precision and the performances of the CRSR. Here, to show that our approach is broadband, we reconstructed the whole spectrum from fig S1. For this reconstruction, we begin with the line at 573 GHz and then added the lines at higher frequency, few by few up to 4 THz (the last observable line is around 3.5 THz) without using data from Hitran except for the 0.38 factor of the doublet. The result of the reconstruction is shown on table S1. It is important to state here that the only input from Hitran remains the 0.38 factor for the 1.21 THz doublet. We did not use any other information than this factor in our whole study and thus to retrieve the following table.

| Upper State | | | Lower State | | | literature values | | Retrieved value | | difference | | Relative difference | |
|---|---|---|---|---|---|---|---|---|---|---|---|---|---|
| J' | K' | a/s | J" | K" | a/s | Freq.(THz) P=0 mbar | Width (MHz) P=10 mbar | Freq. (THz) | Width (MHz) | Frequency (MHz) | Width (MHz) | Freq. | Width |
| 1 | 0 | s | 1 | 0 | a | 0.5725 | 215.395 | 0.57253 | 199.927 | 31.6 | 15.5 | 5.51 10$^{-5}$ | 0.0718 |
| 2 | 1 | s | 1 | 1 | a | 1.1685 | 258.592 | 1.1685 | 303.315 | 36.4 | 44.7 | 3.11 10$^{-5}$ | 0.173 |
| 2 | 0 | a | 1 | 0 | s | 1.2149 | 195.867 | 1.2148 | 219.763 | 51.3 | 23.9 | 4.23 10$^{-5}$ | 0.122 |
| 2 | 1 | a | 1 | 1 | s | 1.2152 | 258.592 | 1.2152 | 218.024 | 28.8 | 40.6 | 2.37 10$^{-5}$ | 0.157 |
| **3** | **0** | **s** | **2** | **0** | **a** | **1.7635** | **178.706** | **1.7621** | **1365.58** | **1410** | **1190** | **8.01 10$^{-4}$** | **6.64** |
| 3 | 1 | s | 2 | 1 | a | 1.7636 | 234.330 | 1.7635 | 113.079 | 109 | 121 | 6.15 10$^{-5}$ | 0.517 |
| 3 | 2 | s | 2 | 2 | a | 1.7638 | 293.505 | 1.7642 | 5.23188 | 373 | 288 | 2.11 10$^{-4}$ | 0.982 |
| 3 | 1 | a | 2 | 1 | s | 1.8089 | 234.330 | 1.809 | 248.148 | 57.7 | 13.8 | 3.19 10$^{-5}$ | 0.059 |
| 3 | 2 | a | 2 | 2 | s | 1.8104 | 293.505 | 1.8104 | 245.363 | 30.2 | 48.1 | 1.67 10$^{-5}$ | 0.164 |
| 4 | 3 | s | 3 | 1 | a | 2.3572 | 213.027 | 2.3574 | 296.553 | 166 | 83.5 | 7.04 10$^{-5}$ | 0.392 |
| 4 | 2 | s | 3 | 2 | a | 2.3577 | 265.101 | 2.3583 | 243.703 | 593 | 21.4 | 2.52E-4 | 0.0807 |
| 4 | 3 | s | 3 | 3 | a | 2.3586 | 320.133 | 2.3588 | 14.1798 | 237 | 306 | 1E-4 | 0.956 |
| **4** | **0** | **a** | **3** | **0** | **s** | **2.4000** | **163.913** | **2.4** | **3** | **3.4** | **161** | **1.42 10$^{-6}$** | **0.982** |
| 4 | 1 | a | 3 | 1 | s | 2.4006 | 213.027 | 2.4004 | 565.219 | 154 | 352 | 6.4 10$^{-5}$ | 1.65 |
| 4 | 2 | a | 3 | 2 | s | 2.4023 | 265.101 | 2.4023 | 155.681 | 29.2 | 109 | 1.21 10$^{-5}$ | 0.413 |
| 4 | 3 | a | 3 | 3 | s | 2.4051 | 320.133 | 2.4051 | 312.653 | 0.42 | 7.48 | 1.75 10$^{-7}$ | 0.0234 |
| 5 | 0 | s | 4 | 0 | a | 2.9484 | 150.894 | 2.9486 | 463.737 | 162 | 313 | 5.5 10$^{-5}$ | 2.07 |
| 5 | 1 | s | 4 | 1 | a | 2.9487 | 193.500 | 2.949 | 543.771 | 261 | 350 | 8.84 10$^{-5}$ | 1.81 |
| 5 | 2 | s | 4 | 2 | a | 2.9495 | 239.064 | 2.9493 | 451.308 | 233 | 212 | 7.89 10$^{-5}$ | 0.888 |
| 5 | 3 | s | 4 | 3 | a | 2.9508 | 287.587 | 2.9508 | 420.287 | 0.28 | 133 | 9.49 10$^{-8}$ | 0.461 |
| 5 | 4 | s | 4 | 4 | a | 2.9526 | 339.069 | 2.9527 | 33.2146 | 106 | 306 | 3.59 10$^{-5}$ | 0.902 |
| 5 | 1 | a | 4 | 1 | s | 2.9896 | 193.500 | 2.9897 | 288.323 | 53.9 | 94.8 | 1.8 10$^{-5}$ | 0.49 |
| 5 | 2 | a | 4 | 2 | s | 2.9916 | 239.065 | 2.9916 | 167.763 | 7.19 | 71.3 | 2.4 10$^{-6}$ | 0.298 |
| 5 | 3 | a | 4 | 3 | s | 2.9948 | 287.587 | 2.9948 | 237.209 | 36.1 | 50.4 | 1.2 10$^{-5}$ | 0.175 |
| 5 | 4 | a | 4 | 4 | s | 2.9994 | 339.069 | 2.9995 | 309.077 | 108 | 30 | 3.59 10$^{-5}$ | 0.0885 |
| 6 | 1 | s | 5 | 1 | a | 3.5374 | 176.931 | 3.5373 | 47.23 | 92.8 | 130 | 2.62 10$^{-5}$ | 0.733 |
| 6 | 2 | s | 5 | 2 | a | 3.5385 | 215.394 | 3.5386 | 60.1745 | 86.7 | 155 | 2.45 10$^{-5}$ | 0.721 |
| 6 | 3 | s | 5 | 3 | a | 3.5403 | 257.408 | 3.5404 | 334.802 | 87.2 | 77.4 | 2.46 10$^{-5}$ | 0.301 |
| 6 | 4 | s | 5 | 4 | a | 3.5428 | 301790 | 3.5429 | 156.65 | 81.7 | 302000 | 2.31 10$^{-5}$ | 0.999 |
| 6 | 5 | s | 5 | 5 | a | 3.5460 | 349.721 | 3.546 | 455.652 | 31.2 | 106 | 8.81 10$^{-6}$ | 0.303 |
| 6 | 0 | a | 5 | 0 | s | 3.5749 | 140.835 | 3.5752 | 353.929 | 262 | 213 | 7.32 10$^{-5}$ | 1.51 |
| 6 | 1 | a | 5 | 1 | s | 3.5756 | 176.932 | 3.5754 | 645.983 | 170 | 469 | 4.74 10$^{-5}$ | 2.65 |
| 6 | 2 | a | 5 | 2 | s | 3.5777 | 215.395 | 3.5778 | 284.482 | 111 | 69.1 | 3.09 10$^{-5}$ | 0.321 |
| 6 | 3 | a | 5 | 3 | s | 3.5813 | 257.409 | 3.5813 | 413.398 | 14.3 | 156 | 4 10$^{-6}$ | 0.606 |
| 6 | 4 | a | 5 | 4 | s | 3.5865 | 301.790 | 3.5865 | 245.971 | 24.8 | 55.8 | 6.91 10$^{-6}$ | 0.185 |
| 6 | 5 | a | 5 | 5 | s | 3.5933 | 349.721 | 3.5933 | 700.185 | 13.6 | 350 | 3.78 10$^{-6}$ | 1 |

**Table S 1 : Value of the retrieved parameter for the whole spectrum at 10 mbar. The multiplets (lines closer than the usual resolution of 1.2GHz) are shown by using a gray background. In red, we show the lines where the retrieval is worse than the resolution or limited by the methods (reaching the lower bound for the linewidth)**

Here one can see that we retrieved the line central frequency and the width better than the resolution for all the lines except one (35 over 36). The second criteria was met at a precision of 50 % for two third of the lines (24 over 36). It shows that the methods stay valid on most of the spectrum. Still a lack of accuracy exist when dealing with a complex multiplet because more information is needed for the same signal to noise ratio as the one around 1.76 THz.

Additionally, when going at high frequency the signal to noise ratio decrease, the precision decrease as well. It shows that the methods is valid on the whole spectrum and clearly benefit from a high signal to noise ratio. Fig S2 illustrate the sensitivity to the J- and K-dependence.

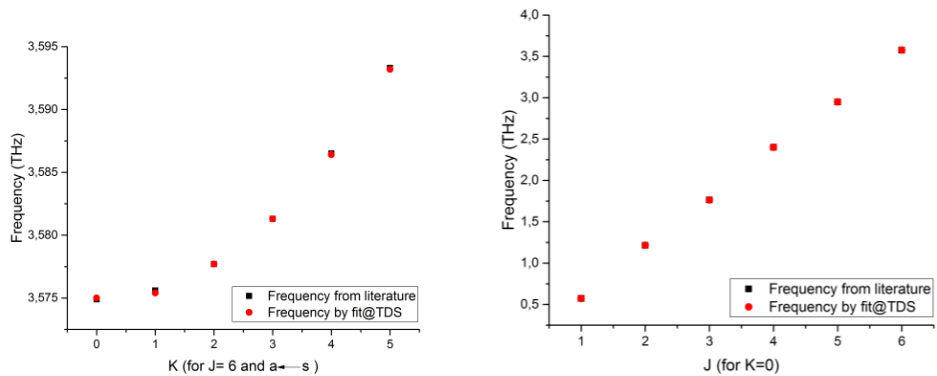

Figure S3: J-dependence (left) and K-dependence (right) for the frequency retrieved by the CRSR method

Overall, it shows that the approach is fully broadband and only limited by the spectroscopic set up to the signal analysis.

Group of 4 lines around 3 THz

In the main text, we showed the performances of the CRSR on lines where the TDS spectrum has most energy. To show that our approach is applicable on the whole spectrum and that the performances will only depend on the signal to noise ratio, we show how the CRSR can retrieve the parameters from the lines around 3 THz. The results are plotted on fig S2 comparable to fig 3 of the main text.

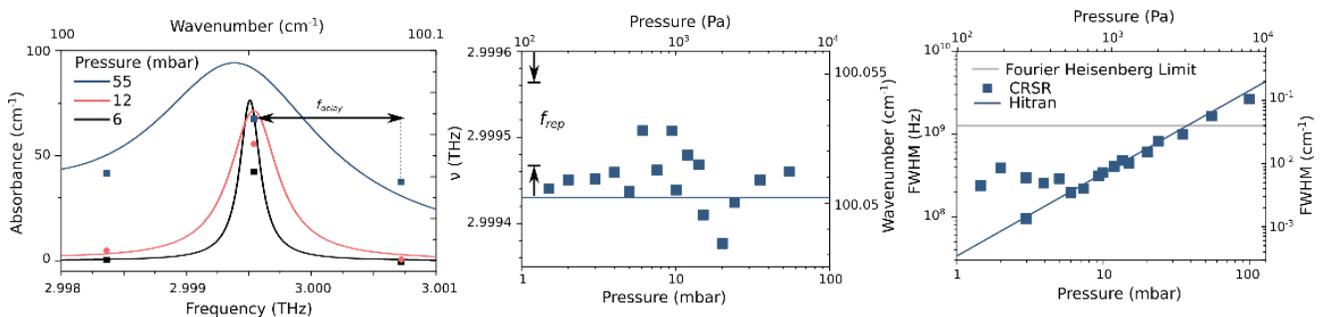

Figure S 4: Super resolution for NH3: results and performances assessment, additional data around 3 THz. Panel A, shows the reconstructed 2.999 THz lines in plain line superimposed to the usual FFT spectrum demonstrating the super resolution reconstruction. Panel B illustrates that the frequency of the line is precisely retrieved (±100 MHz) for pressure from 2 mbar to 100 mbar. Panel C shows the linewidth of the line versus the pressure demonstrating precise retrieval for pressure from 6 mbar to 100 mbar.

Here one can see that we are able to reach performance comparable to the one obtained at higher energy and conclude that the methods is valid on the broad THz-TDS spectrum.

4. **Absolute frequency normalization on water lines**

In a TDS system, the frequency quantification and its precision comes from quantification and the precision of the time sampling step that is by itself tainted with error [2]. The manufacturer corrected these measurements for systematics error to reduce it to the minimum and to get an as good as possible absolute value of the time delay. However, this cannot be done at the precision needed for super resolution, simply because CRSR is not implemented yet on the commercial system. To compensate for this, we recorded the spectrum of water vapor around 570 GHz, and we used the CRSR to get the central frequency of this very known line. We used the ratio found (1.00086423) between the database value and the retrieved value to normalize the value of our time step and get better absolute measurement. We estimated the precision on the time step to be better than 4 10-5 from the precision on the NH3 lines central frequency.

5. **Stability of the repetition frequency of the fiber laser**

We measured the stability of repetition rate of the laser to be sure this will not be a limitation for our experiments. We found a long time deviation of 30 Hz on the repetition rate that is relative a precision of 3 10$^{-7}$, or a variation in repetition time of 3 fs. This is in good agreement with the 1.405 fs of the delay noise of two consecutive experiments (see noise evaluation). It is important to note that this jitter is below 10 % of the time step (100/3 of fs). It means that for narrow line with time spanning wider than the repetition time, the measurement will still be in the good time step. Said differently, one can sample the signal originate from one pulse of the laser with the next pulse and still be in the good time step which is an important prerequisite if we want to reach super resolution better than the repetition frequency.

6. **Noise evaluation**

To better analyze our data we first perform a rapid noise evaluation. Figure S5 shows the reference spectrum, together with the difference between two consecutive references, and the difference between two dark noises (time traces recorded when the THz path is fully blocked). It has to be noted that the time shift correction will automatically be taken into account by the refractive index term in the CRSR.

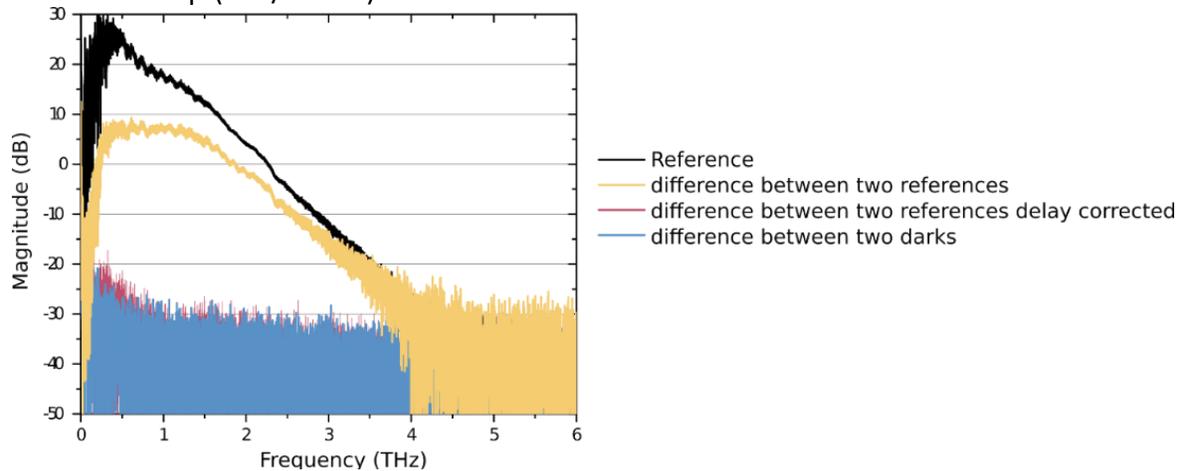

Figure S5 Reference Signal superimposed to the difference between two reference signals (corrected from a time shift) and with the difference between two spectrum where the THz signal is blocked (dark noises).

Here, one can see that the noise comes from the time delay, the dark noise variation and an additional noise terms. To figure out from where this noise comes from, the noise (the "difference" curves) are plotted versus the signal of a reference (panel A) and the derivative of the signal of a reference (panel B) on Figure S 6.

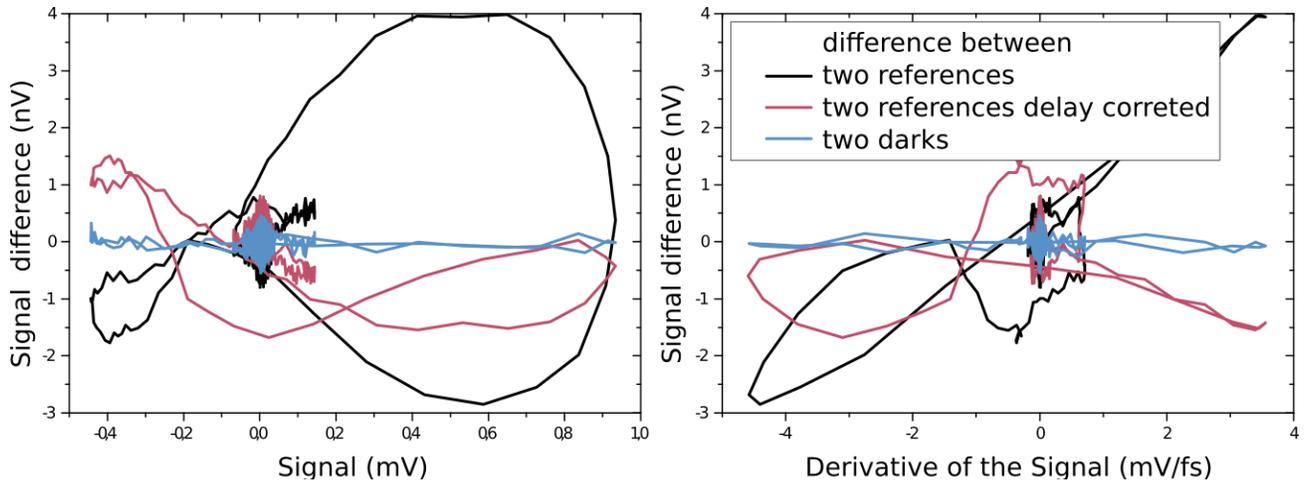

Figure S 6 : Noise (difference between two consecutive signals) versus signal A, and derivative of the signal B both compared with and without fixed time shift correction and compared to the dark noise. The x-axis is always taken on the reference signal data even when plotting the dark noise.

Here we can see that there is a correlation of the highest discrepancy between the two signals and the signal value. However, one cannot claim proportionality or a quadratic law between them. To go a step further looking the signal difference versus the derivative of the signal, one can clearly see that the error is proportional to the derivative of the signal. This means that this error corresponds to a fixed minor delay between the two time traces. To verify this hypotheses we simply run an optimization on the time delay between those curves regarding the difference between its. We found a time delay of 1.405 fs about 1/20 of the time sampling. We attributed this time delay to the routine used by the TDS to correct the time sampling variation due to the delay line. At the end, quantitatively, the difference of the two signals account for $10^{-4}$ of the whole signal (in energy) and the difference of the two darks for $3\ 10^{-5}$. This means that even if additional correction on the signal could be done, we could not improve the signal to noise ratio by more than a factor of three and we consequently decided to stop here.

7. **Other supportive information**

The reader will find the source code for the super resolution, the one to create the numerical examples, the numerical examples we created as the data from the NH3 experiment in zip files linked to this paper.